\documentclass[aps,prl,twocolumn]{revtex4}
\usepackage{graphicx}

\usepackage{bm}
\usepackage{color}

\newcommand{\ke}[1]{|#1\rangle}

\begin{document}

\title{Cavity cooling of internal molecular motion}

\author{Giovanna Morigi,$^1$ Pepijn W.H. Pinkse,$^2$ Markus
Kowalewski,$^3$ and Regina de Vivie-Riedle $^3$}

\affiliation{$^1$ Departament de Fisica, Universitat Autonoma de
Barcelona,
E-08193 Bellaterra, Spain,\\
$^2$ Max-Planck-Institut f\"ur Quantenoptik, D-85748 Garching,
Germany,\\
$^3$ Departement of Quantum Chemistry,
Ludwig-Maximilian-Universit\"at M\"unchen, D-81377 M\"unchen,
Germany}

\date{\today} \begin{abstract} We predict that it is possible to
cool rotational, vibrational and translational degrees of freedom
of molecules by coupling a molecular dipole transition to an
optical cavity. The dynamics is numerically simulated for a
realistic set of experimental parameters using OH molecules. The
results show that the translational motion is cooled to few $\mu$K
and the internal state is prepared in one of the two ground states
of the two decoupled rotational ladders in few seconds. Shorter
cooling times are expected for molecules with larger
polarizability. \end{abstract}

\maketitle


The preparation of molecular samples at ultra-low temperatures
offers exciting perspectives in physics and
chemistry~\cite{quovadis}. This goal is presently pursued by
several groups worldwide with various approaches. Two methods for
generating ultracold molecules employ photoassociation and
Feshbach resonances, and are efficiently implemented on
alkali dimers~\cite{quovadis}. Another approach uses buffer gases,
to which the molecules thermalize~\cite{Weinstein98}. Its
application is limited by the physical properties of atom-molecule
collisions at low temperatures. Optical cooling of molecules is an
interesting alternative, but, contrary to atoms, its efficiency is
severely limited by the multiple scattering channels coupled by
spontaneous emission, and may only be feasible for molecules which
are confined in external traps for very long
times~\cite{Drewsen04}. Elegant laser-cooling proposals, based on
optical pumping the ro-vibrational states~\cite{Stwalley,Drewsen}
and excitation pulses tailored with optimal control
theory~\cite{Tannor,Q-Control-Molecules}, exhibit efficiencies
which are indeed severely limited by spontaneous decay.
In~\cite{Horak,Vuletic00} it was argued that cooling of the
molecular external motion could be achieved by using resonators,
by enhancing stimulated photon emission into the cavity mode over
spontaneous decay. This mechanism was successfully applied for
cooling the motion of atoms~\cite{Maunz04}.

\begin{figure}[ht]
\includegraphics[width=0.35\textheight]{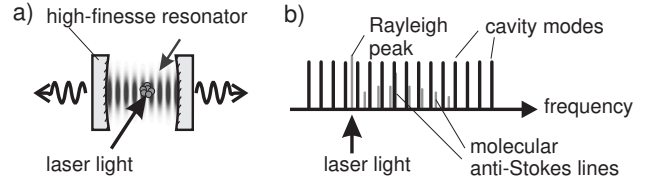}%
\caption{(a) A molecular sample interacts with the cavity field
and is driven by a laser, inducing Raman transitions cooling the
internal and external degrees of freedom. (b) Comb of resonances
at which photon emission into the cavity is enhanced. The grey
bars symbolize the molecular lines, which in OH extend over
several tens nm. The laser frequency (arrow) is varied to
sequentially address several anti-Stokes lines.}
\label{Fig:1}%
\end{figure}

In this Letter we propose a method for optically cooling external
as well as the {\it internal} degrees of freedom of molecules. The
method relies on the enhancement of the anti-Stokes Raman
transitions through the resonant coupling with the modes of a
high-finesse resonator, as sketched in Fig.~\ref{Fig:1}. All
relevant anti-Stokes transitions are driven by sequential tuning
of the driving laser. At the end of the process the molecule is in
the ro-vibrational ground state and the motion is cooled to the
cavity linewidth. We demonstrate the method with {\it ab initio}
based numerical simulations using OH radicals, of which cold
ensembles are experimentally
produced~\cite{vdMeerakker05,Bochinski04}.

We now outline the theoretical considerations.  We consider a gas
of molecules of mass $M$, prepared in the electronic ground state
$X$, and with dipole transitions $X\to {E}$. Here, ${E}$ is a set
of electronically excited states, including higher-lying states
which may contribute significantly to the total polarizability. We
denote the ro-vibrational states and their corresponding
frequencies by $|j,\xi\rangle$ and $\omega_{\xi}^{(j)}$
($j=X,A,...$), while the elements of the dipole moment ${\bf d}$
are ${\cal D}_{\xi\to\xi''}=\langle A \in {E},\xi''|{\bf
d}|X,\xi\rangle$. These transitions are driven by a far-off
resonant laser and interact with an optical resonator as
illustrated in Fig.~\ref{Fig:1}(a). In absence of the resonator,
spontaneous Raman scattering determines the relevant dynamics of
molecule-photon interactions. These processes depend on the
center-of-mass momentum ${\bf p}$ through the Doppler effect and
occur at rate $\Gamma_{\xi\to\xi'}^{(\gamma)}({\bf
p})=\sum_{E,\xi''}\Gamma_{\xi''\to\xi'}
   \gamma_{E,\xi\to\xi''}({\bf
p}) $, where $\Gamma_{\xi'' \to
\xi'}$ the decay rate along the transition $|A,\xi''\rangle\to|X,\xi'\rangle$, and
\begin{equation}
 \gamma_{E,\xi\to\xi''}({\bf p})=\frac{
    \Omega_{L,\xi\to\xi''}^2}{(\Delta_{\xi,\xi''}^E+{\bf k_L}\cdot {\bf
    p}/M)^2+\Gamma_{\xi''}^2/4},
\label{eq:Gammag} \end{equation} with
$\Gamma_{\xi''}=\sum_{\xi'}\Gamma_{\xi''\to\xi'}$. Here,
$\Omega_{L,\xi\to\xi''}={\cal E}_L({\cal D}_{\xi\to\xi''}\cdot
\epsilon_L)/\hbar$ gives the strength of coupling to the laser
with electric field amplitude ${\cal E}_L$, polarization
$\epsilon_L$, frequency $\omega_L$, and wavevector ${\bf k_L}$,
and
$\Delta_{\xi,\xi''}^E=\omega_{\xi}^{(X)}-\omega_{\xi''}^{(E)}+\omega_L$
denotes the detuning between laser and internal transition. In the
limit where the laser is far-off resonant from the electric dipole
transitions, electronic ground states at different ro-vibrational
quantum numbers are coupled via Raman transitions. The
corresponding emission spectrum is symbolized by the grey bars in
Fig.~\ref{Fig:1}(b). In this regime, coupling an optical resonator
to the molecule may enhance one or more scattering processes when
the corresponding molecular transitions are resonant with
resonator modes, symbolized by the black bars in
Fig.~\ref{Fig:1}(b). This enhancement requires that the rate
$\Gamma_{\xi\to\xi'}^{\kappa}({\bf p})$, describing scattering of
a photon from the laser into the cavity mode and its subsequent
loss from the cavity, exceeds the corresponding spontaneous Raman
scattering rate, $\Gamma_{\xi\to\xi'}^{\kappa}({\bf
p})\gg\Gamma_{\xi\to\xi'}^{\gamma}({\bf p})$. For a standing wave
cavity, in the regime where the molecular kinetic energy exceeds
the cavity potential, and when the cavity photon is not reabsorbed
but lost via cavity decay~\cite{Vuletic00}, we have
$\Gamma_{\xi\to\xi'}^{\kappa}({\bf
p})=\Gamma_{\xi\to\xi'}^{\kappa,+}({\bf
p})+\Gamma_{\xi\to\xi'}^{\kappa,-}({\bf p})$, where the sign $\pm$
gives the direction of emission along the cavity axis and
\begin{eqnarray}
    \Gamma_{\xi\to\xi'}^{\kappa,\pm}({\bf p})
    &=&2\kappa\sum_{c,E,\xi''}\frac{\gamma_{E,\xi\to\xi''}({\bf
p})\Bigl|g_{c,\xi''\to\xi'}^{\pm}\Bigr|^2 }
    {\left(\delta\omega\pm
    {\bf k_c}\cdot {\bf p}/M\right)^2+\kappa^2}.
    \label{eq:GammaK}
\end{eqnarray} Here, $2\kappa$ is the cavity linewidth,
$\delta\omega=\omega_{\xi}^{(X)}-\omega_{\xi'}^{(X)}+\omega_L-\Omega_c$
is the frequency difference between initial and final (internal
and cavity) state, with $\Omega_c$ the frequencies of the cavity
modes, and $g_{c,\xi'\to\xi''}^{\pm}$ are the Fourier components
at cavity-mode wave vector $\pm |{\bf k_c}|$ of the coupling
strength to the empty cavity mode, $g_{c,\xi'\to\xi''}({\bf
x})={\cal E}_c({\bf x})(\epsilon_0\cdot{\cal
D}_{\xi',\xi''})/\hbar$, with ${\cal E}_c$ and $\epsilon_0$ vacuum
amplitude and polarization. Note that reabsorption and spontaneous
emission of the cavity photon, can be neglected while $\kappa\gg
|g_{c,\xi'\to\xi''}\Omega_{L,\xi\to\xi''}/\Delta_{\xi,\xi''}^E|$.

Enhancement of {\it Rayleigh} scattering into the cavity is
achieved by setting the laser on resonance with one cavity mode,
$\omega_L=\Omega_c$, see Fig.~\ref{Fig:1}(b), provided that
$\Gamma_{\xi\to\xi}^{(\gamma)}\ll\Gamma_{\xi\to\xi}^{(\kappa)}$,
i.e., $g_{c,\xi\to\xi}^2/\Gamma\kappa\gg 1$, where $\Gamma$ is
determined by the linewidths of the excited states which
significantly contribute to the scattering process. This situation
has been discussed in~\cite{Vuletic00}, where it has been
predicted that the motion can be cavity cooled to a temperature
which is in principle only limited by the cavity
linewidth~\cite{DomokosRitschJOSA03}, provided that the laser is
set on the low-frequency side of the cavity resonance. Note that
cooling of the motion in the plane orthogonal to the cavity axis
is warranted when the laser is a standing wave field, which is
simply found in our model by allowing for the absorption of laser
photons at wave vector $-{\bf k_L}$. In general, enhancement of
scattering along the Raman transition $\xi\to\xi^{\prime}$,
decreasing the energy of the ro-vibrational degrees of freedom, is
achieved by setting the laser such that the corresponding
anti-Stokes spectral line is resonant with one cavity mode, and
requires $g_{c,\xi\to\xi'}^2/\Gamma\kappa\gg 1$. The cooling
strategy then consists of choosing a suitable cavity and of
sequentially changing the laser frequency, so to maximize the
resonant drive of the different anti-Stokes spectral lines, and
thereby cooling the molecule to the ro-vibrational ground state.

\begin{figure}[t]
\includegraphics[width=0.3\textheight]{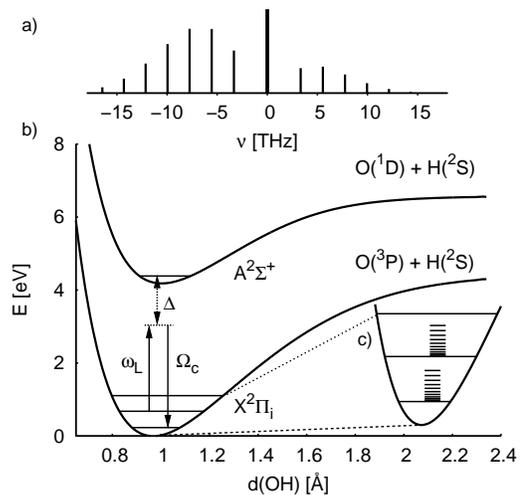}
\caption{(a) Simulated Raman spectra for the first nine rotational
states of OH, which are relevantly occupied at room temperature.
(b) Potential energy surfaces of the X$^2\Pi_i$ ground state and of
the A$^2\Sigma^+$ excited state. The coherent Raman process is
indicated by the arrows. (c) Ro-vibrational substructure.}
\label{Fig:PES}%
\end{figure}

We simulate the cooling dynamics for OH radicals using a rate
equation based on the rates~(\ref{eq:Gammag})
and~(\ref{eq:GammaK}). The considered Raman process is detuned
from the excitation energy of $32402\,$cm$^{-1}$ ($971.4$~THz)
between the $X^2\Pi_i$ ground state and the electronically excited
state $A^2\Sigma^+$ as indicated in Fig.~\ref{Fig:PES}(b). The
relevant ro-vibrational spectrum and the coupling of the molecule
to the laser field and cavity modes were obtained by combining
{\it ab initio} calculations for the vibronic degrees of freedom
with available experimental data for the rotational constants. The
level scheme is displayed in Fig.~\ref{Fig:PES}(b). The potential
energy surfaces (PES) and the polarizabitilies were calculated
with highly correlated quantum chemical methods: the electronic
structure calculations \cite{molpro} were performed on the Multi
Configurational SCF level using a single atom basis set
(aug-cc-pVTZ). Rates (\ref{eq:Gammag}) and (\ref{eq:GammaK}) were
evaluated using the polarizability tensors, defined as
$\alpha_{\xi\to\xi'}=\sum_{\xi''} {\cal D}_{\xi \to \xi''} {\cal
D}_{\xi'' \to \xi'} / \hbar
\Delta_{\xi''}$~\cite{Footnote:Polarizability} and calculated with
linear response theory at the MCSCF level~\cite{linres1, linres2,
dalton}. In order to determine the internal level structure and
transition strengths, the vibrational eigenvalues and
eigenfunctions were evaluated with a relaxation method using
propagation in imaginary time plus an additional diagonalization
step~\cite{Sundermann}. The corresponding Placzek-Teller
coefficients~\cite{placzekteller} were calculated for the
transitions between the rotational sublevels.

In order to obtain a concise picture of the cooling dynamics,
several assumptions were made without loss of generality. The
molecules are prepared in the lower-lying $X^2\Pi_{3/2}$ component
of the $X^2\Pi$ electronic ground state. The hyperfine splitting
is neglected as angular momentum conservation for the rotational
Raman transitions inhibits transitions between the hyperfine
sublevels. At $T\approx 300$~K only the first 9 rotational states
of OH are relevantly occupied, while only the vibrational ground
state is populated. The selection rule $\Delta J = 0,~\pm 2$ for
the rotational transitions yields that the matrix elements of
transitions between rotational states with opposite parity vanish,
resulting in two separate ladders for the scattering processes
with final states $\ke{X,v=0,J=0,1}$~\cite{Footnote:Spectra}. We
assume that a preparation step has occurred, bringing the motional
temperature to below~$1\,$K. This could be realized with, e.g.,
helium-buffer-gas cooling~\cite{Weinstein98}, electrostatic
filtering~\cite{RangwalaPRA03}, or decelerator
techniques~\cite{BethlemPRL99,Barker}. The high-finesse optical
cavity has a free-spectral range (FSR) of $15\times2\pi\,$GHz,
which can be realized with a Fabry-Perot-type cavity of length
$L=1\,$cm. For simplicity we assume that the cavity only supports
zeroth-order transverse modes. This actually underestimates the
possibilities of scattering light into the cavity, since
higher-order transverse modes can be combined in degenerate
cavities, like confocal resonators. The cavity half-linewidth is
set to $\kappa=75\times2\pi\,$kHz, and the coupling $g_{c,0\to 0}
= 2 \pi\times 116$ kHz. This is achieved with a mode volume of
$3.2 \times 10^{-13}~$m$^{3}$, assuming a mode waist of $w_0 =
6~\mu$m, and a cavity finesse $F=10^5$, i.e., a mirror
reflectivity of $0.999969$. We also choose a laser wavelength of
532\,nm, for which ample power is available as well as mirrors of
the required quality. The frequency of the laser is far below that
of the OH A-X ro-vibronic band. We assume to have single-frequency
light of 10\,W enhanced by a factor of 100 by a build-up cavity in
a TEM$_{00}$ mode, corresponding to a Rabi coupling
$\Omega_{L,0\to 0}=2 \pi\times 69\,$GHz and frequency
$\omega_L=\omega_0^{(A)}-\omega_0^{(X)}-\Delta$ with
$\Delta\approx 2\pi\times407\,$THz.
\begin{figure}[htb]
\includegraphics[width=0.3\textheight]{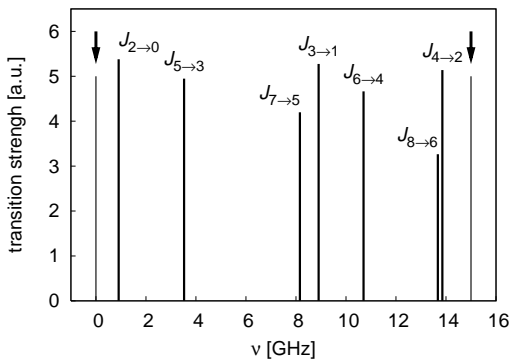}
\caption{Reduced spectrum of the relevant rotational anti-Stokes
transitions: the lines are projected onto a single free-spectral
range of the cavity, whose width is indicated by the arrows. The
unfolded Raman spectrum spans 15\,THz or up to $10^3$ cavity
modes. The frequencies of the lines are evaluated from quantum
chemical calculations, and must be understood as a qualitative
picture; high-resolution experimental input is needed to fix the
absolute position with kHz accuracy.} \label{Fig:modFSR}
\end{figure}
The latter value is sequentially varied during cooling, in order
to drive (quasi-)resonantly the cooling transitions. In
combination with the broad spectrum of cavity modes, the laser
only needs to be varied over one FSR to address all anti-Stokes
lines. Fig.~\ref{Fig:modFSR} displays the anti-Stokes Raman lines
as a function of the frequency modulus the FSR. Addressing the
Stokes lines (not drawn) can be avoided, given the small laser and
cavity linewidths. 
In a confocal cavity, e.g., all higher-order transverse cavity
modes will be degenerate with fundamental ones.
In addition, our scheme is robust against a small number of
coincidences between Stokes and anti-Stokes lines.

\begin{figure}
\includegraphics[width=0.35\textheight]{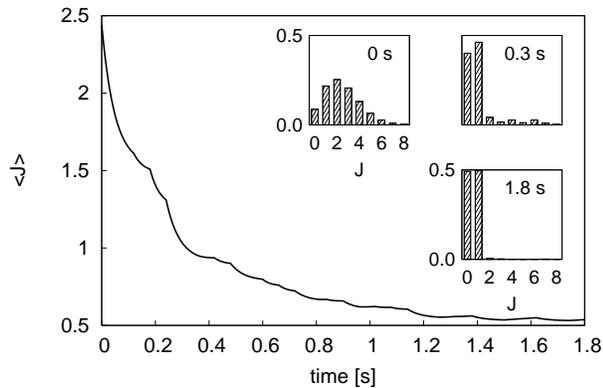}
\caption{Mean rotational quantum number versus time during the
cooling process. The manually optimized cooling sequence first
empties the levels $J=(2,3)$, thereafter higher levels are
addressed. The small figures show the state distributions at time
0, 0.3 and 1.8\,s. The cavity length is fine-tuned to address the
$J_{2\rightarrow0}$ and the $J_{3\rightarrow1}$ transition
simultaneously.} \label{Fig:meanJ} \end{figure}

The cooling strategy is as follows. First, the external degrees of
freedom are cooled to the cavity linewidth, corresponding to a
final temperature $T\approx 4\mu$K, by setting the Rayleigh
transition quasi-resonant with one cavity mode. The corresponding
coefficients have been evaluated numerically, giving the rate of
Rayleigh scattering for OH into the cavity $\Gamma^{\kappa}_{0\to
0}\sim 1$ kHz, while the spontaneous rate
$\Gamma^{\gamma}_{\xi\to\xi}\sim 0.5-2.5$ Hz. We verified the
efficiency of cooling by solving the semiclassical equation for
the mechanical energy~\cite{Itano}. For these parameters, starting
from $T\sim 1$K for the external degrees of freedom, the cooling
limit is reached in a time of the order of $1$~s. Then, the
rotational degrees of freedom are cooled by setting the laser
frequency to sequentially address each anti-Stokes spectral line.
A manually optimized sequence led to the result in
Fig.~\ref{Fig:meanJ}, where the mean rotational quantum number
$\langle J \rangle$ is plotted as a function of time. The final
value $\langle J\rangle\approx 0.5$ corresponds to the final
situation in which the two states $J=0$ and $J=1$, ground states
of each ladder, achieve maximum occupation, equal to $50\%$. The
insets in Fig.~\ref{Fig:meanJ} show that after $0.3$~s their
occupation is about 40\%, while after $1.8$~s it reaches 49\%
(leading to a total population of 98.8\%). The cooling rate for
the rotational degrees of freedom of OH is of the order of $4$~Hz,
see Fig.~\ref{Fig:meanJ}, while the rate of heating due to
spontaneous Raman scattering along the Stokes transitions is about
$0.1$~Hz. In the simulation the vibrational degrees of freedom are
taken into account but no vibrational heating is observed. In a
separate simulation, we checked that vibrational excitations are
cooled with the same scheme.

The cooling time can be improved significantly for molecules with
higher polarizabilities $\alpha$, as the cooling rate scales with
$\alpha^2$. For instance, the polarizability of NO is
approximately 4 times larger than for OH, and preliminary results
show that it indeed cools down faster, although the
cooling time is not reduced by a factor 16 because more
rotational states are occupied at $300\,$K. For molecules like
Cs$_2$, $\alpha$ is two orders of magnitude larger than for OH,
and should yield faster cooling rates.

For molecules with low polarizability, like OH, reduction of the
cooling time seems possible by further optimizing the sequential
procedure. In addition, the efficiency could be improved by using
degenerate cavity modes or by superradiant enhancement of light
scattering, sustained by the formation of self-organized molecular
crystals~\cite{CollectiveCooling}.

In summary, we presented a strategy for cooling external and
internal degrees of freedom of a molecule. We simulated the
cooling dynamics for OH using experimentally accessible parameter
regimes, showing that this method allows for efficient preparation
in the lowest rovibrational states, while the motion is cooled to
the cavity linewidth. For OH the cooling time is of the order of
seconds, and requires thus the support of trapping technologies
which are stable over these
times~\cite{Weinstein98,BethlemNature00,RiegerPRL05,DeMilleEPJD04,Drewsen04}.
The cooling time of molecules with larger polarizabilities can
scale down to few ms, when the polarizability is about 10 times
larger. Applications of this technique to polyatomic molecules has
to deal with an increasing number of transitions to be addressed,
which will slow down the process. A possible extension of this
scheme could make use of excitation pulses, determined with
optimal control techniques~\cite{Tannor}.

\begin{acknowledgments} G.M. thanks the Theoretical Femtochemistry
Group at LMU for the hospitality. Support by the European
Commission (CONQUEST, MRTN-CT-2003-505089; EMALI,
MRTN-CT-2006-035369), the Spanish MEC (Ramon-y-Cajal; Consolider
Ingenio 2010 "QOIT"; HA2005-0001), EUROQUAM (Cavity-Mediated
Molecular Cooling), and the DFG cluster of excellence Munich
Centre for Advanced Photonics, is acknowledged.
\end{acknowledgments}

\end{document}